\documentclass[11pt,twoside]{article}

\usepackage{asp2006}
\usepackage{graphicx}
\usepackage{lscape}

\bibpunct{(}{)}{,}{a}{}{,}

\begin{document}
\title{VLBI Surveys of Active Galactic Nuclei}
\author{Y.~Y.~Kovalev}
\affil{
Max-Planck-Institut f\"ur Radioastronomie,
Auf dem H\"ugel\,69, 53121~Bonn, Germany;
\\
Astro Space Center of Lebedev Physical Institute,
Profsoyuznaya 84/32, 117997 Moscow, Russia
} 

\begin{abstract}
A review is given on the current status and selected results from large
VLBI surveys of compact extragalactic radio sources made
between 13~cm and 3~mm wavelengths and covering the entire sky. More
than 4200 objects are observed and imaged with dynamic ranges from
a hundred to several thousand at (sub)parsec scales.
Implications to the VSOP-2 project are discussed.
\end{abstract}

\section{Introduction}

During the last several decades, a number of large Very Long Baseline
Interferometry (VLBI) surveys were
conducted covering the frequency range between 2 and 100~GHz including
the first space VLBI survey with the space radio telescope HALCA
\citep[e.g.,][]{Lovell_etal04,Dodson_VSOPsurvey}. VLBI surveys which
have observed and produced images for more then 100 objects are
summarized in Table~\ref{t:VLBIsurveys}. We discuss below selected
results from these surveys as well as their implications to the planned
next generation science driven space VLBI mission VSOP-2
\citep{VSOP2_2004}.

\section{Selected results from large VLBI surveys of extragalactic jets}
\label{s:survey_results}

\begin{table}[t!]
\caption{Large VLBI surveys of extragalactic radio sources.
\label{t:VLBIsurveys}
}
\smallskip
\begin{center}
{
\footnotesize
\begin{tabular}{lcrll}
\tableline
\noalign{\smallskip}
Name & $\lambda$ & No.~of    & Recent    & Comments \\
     & (cm)      & sources   & reference &          \\
\noalign{\smallskip}
\tableline
\noalign{\smallskip}
CJF survey             & 18 \& 6      & 293           & \cite{PTZ03}            & complete       \\
ICRF/RDV               & 13 \& 3.6    & 500           & \cite{Ojha_etal04}      & open           \\
VLBA Calibrator survey & 13 \& 3.6    & $>3400$       & \cite{vcs5}             & open, complete \\
VSOP VLBApls           & 6            & 374           & \cite{VLBApls}          &                \\
VSOP survey            & 6            & $\approx 300$ & \cite{Dodson_VSOPsurvey}&                \\
VIPS                   & 6            & 1127          & \cite{Hel07}            & open, complete \\
2 cm Survey /          & 2            & 250           & \cite{2cmPaperIV}       & open, complete \\
MOJAVE                 & 2            & $>133$        & \cite{LH05}             & open, complete \\
VERA FSS / VLBA GaPS   & 1.35         & $>500$        & \cite{Petrov_etal07}    & open           \\
ICRFext 22 \& 43 GHz   & 1.35 \& 0.7  & $>100$        & \cite{Lanyi_etal05}     &                \\
GMVA 3mm               & 0.3          & 121           & \cite{Lee08}            &                \\
\noalign{\smallskip}
\tableline
\end{tabular}
}
\end{center}
Note. --- Comments in the last column mean the following.
`Complete': the sample or its sub-sample is complete, flux-density-limited;
`open': all or some of the data are publicly available in a form of $uv$ and/or image FITS files.
\end{table}

\begin{figure}[h!]
\begin{center}
\resizebox{0.6\hsize}{!}{
   \includegraphics[trim = 0cm 0.8cm 0cm 0cm,angle=0]{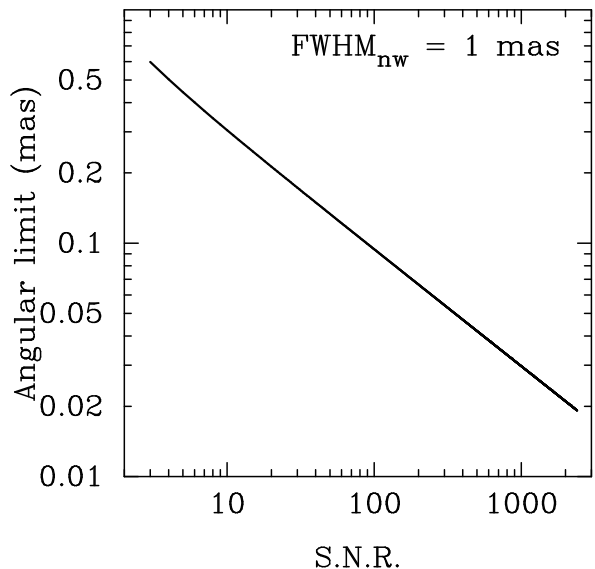}
}
\end{center}
\caption{\label{f:ang_limit}
Dependence of the resolution limit versus signal-to-noise ratio. Results
of theoretical estimation for a circular Gaussian beam with 
the full width at half maximum equal to 1~mas for a naturally weighted image.
}
\end{figure}

Only a few of the sources have an
overall radio structure which is almost unresolved even at the longest
spacings. However, most of the observations do not have enough
resolution to resolve structure across the jet
\citep[e.g.,][]{2cmPaperIV,Hel07}. The distribution of the measured
brightness temperatures of the jet cores, calculated in the source
frame, peaks around $10^{12}$~K and extends up to about $10^{14}$~K
\citep{Horiuchi_etal04,2cmPaperIV,Dodson_VSOPsurvey}. This is
close to the limit set by the dimensions of individual VLBI arrays. However,
for many sources only a lower limit is determined. The observed values
can be explained as the result of Doppler boosting.

It is important to note that the highest measurable value of the
brightness temperature does not depend on the observing frequency
\citep[see equation~(5) in][]{2cmPaperIV}. It depends on the physical
length of the interferometric baselines as well as on the
signal-to-noise ratio (S.N.R.) of VLBI measurements. This is
demonstrated in Figure~\ref{f:ang_limit} as a dependence of the
resolution limit versus S.N.R.\ calculated according to a criteria
proposed by \cite{2cmPaperIV}, equations~(1,2). See for more discussion
also \cite{Lob05}.

Fractional linear polarization of the opaque quasar cores is generally
low ($<5$\,\%). Extended jet features often show strong linear
polarization (up to $\sim 50$~per~cent) indicating synchrotron radiation
of optically thin regions with highly ordered magnetic field
\citep[e.g.,][]{LH05,Hel07}. Circular polarization, if any, is detected,
in general, on the level less than 0.5\% mostly in the cores
\citep{HL06}.

\section{Implications, prospects, and challenges of the VSOP-2 project}

A VSOP-2 experiment to measure brightness temperature of jet cores in a
complete sample of AGNs could be valuable for statistical studies of the
properties of extragalactic jets, for a population analysis of the jet
orientation and other basic parameters like the Doppler factor \cite[see,
e.g.,][]{Lobanov_etal2000,Homan_etal06}. It should be taken into
account, while planning such a survey, that the highest measurable
brightness temperature does not depend on the wavelength of
observations. It depends in turn on the interferometer baseline length
and the S.N.R.\ (see \S~\ref{s:survey_results}). An obvious conclusion
is that 8~GHz observations involving VSOP-2 space telescope and big
ground-based dishes (e.g., EVN) will provide the highest sensitivity of
the ground-space correlated-flux-density measurement being the most
efficient configuration for such a survey.

\begin{figure}[b!]
\begin{center}
\resizebox{\hsize}{!}{
   \includegraphics[trim = 2.1cm 4cm 2.2cm 4cm,clip,angle=270]{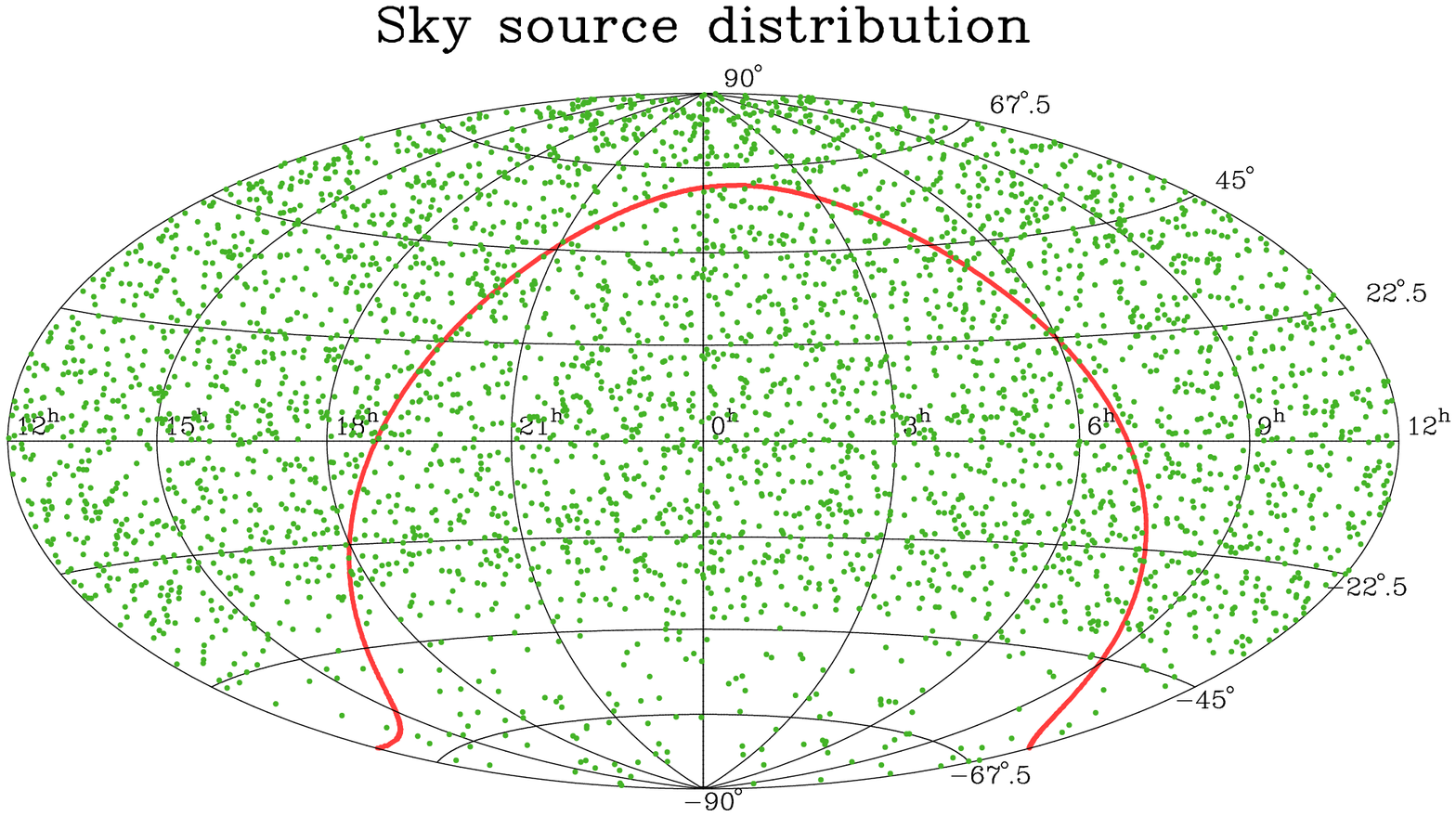}
}
\end{center}
\caption{\label{f:skydist}
Extragalactic radio sources detected in the ICRF/VCS surveys at 2
and/or 8~GHz \citep{icrf98,icrf-ext2-2004,vcs1,vcs2,Ojha_etal04,vcs3,vcs4,vcs5,vcs6}.
}
\end{figure}

It is planned to implement a fast switching capability for the VSOP-2
space radio telescope to allow phase referencing technique to be used
for successful VLBI observations of weak targets. This is especially
important at high radio frequencies which have a very limited coherence
time of less then several minutes. We present here accumulated results
of the search for compact extragalactic sources which are suitable as
phase reference calibrators in VLBI experiments at or below 8~GHz
\citep{icrf98,icrf-ext2-2004,vcs1,vcs2,Ojha_etal04,vcs3,vcs4,vcs5,vcs6}.
Figure~\ref{f:skydist} shows the sky coverage for more than 3000
calibrators while Table~\ref{t:calprob} provides a probability to find a
calibrator within a given search radius. These results incorporate
additional objects found in VLBA surveys recently which were not
included in the \cite{Asaki_etal07} analysis. 
It can be immediately seen that a probability to find a phase calibrator
required for VSOP-2 observations at X-band (8~GHz) within several
degrees of the search radius is quite high for declination
$\delta>-40^\circ$.

It is important to note that the full list of calibrators includes a
complete sub-sample of sources with integrated VLBA flux density greater
than 200\,mJy at 8~GHz and declination $\delta>-30^\circ$ \citep[see for
details][]{vcs5}. This means that there are almost no bright compact
sources ``left'' in that region of the sky.
In order to increase the calibrator source coverage in the declination
range $-50^\circ<\delta<-30^\circ$ a dedicated VLBA survey is proposed.
The calibrator coverage south of declination $-50^\circ$ will be
increased by the Australian LBA survey started in February 2008, this
campaign will observe $\sim 500$ candidate sources (P.I.: L.~Petrov).

The situation at 22 and 43~GHz is not so well advanced currently. Only
several hundred extragalactic objects were observed and imaged in the
framework of the astrometric/geodetic program by \cite{Lanyi_etal05}. In
addition to that, results of dedicated 22~GHz VLBA survey of
extragalactic sources with low galactic latitude will become available
soon. However, 22 and 43~GHz observations of the complete sample of
compact objects \citep{vcs5} are needed in order to find more sources to
be used as VLBI phase calibrators at these high radio frequencies.


\begin{table}[t!]
\caption{Probability to find a calibrator within a given search radius
for two declination ($\delta$) zones following results from
2 and 8~GHz VLBI surveys,
see references in the caption to Figure~\ref{f:skydist}.
Sources were assigned the calibrator classification according to criteria
described in \cite{vcs5}.
\label{t:calprob}
}
\smallskip
\begin{center}
{
\footnotesize
\begin{tabular}{crr}
\tableline
\noalign{\smallskip}
Search radius &\multicolumn{2}{c}{Probability (\%)}\\
(deg)&$\delta>-40^\circ$&$\delta<-45^\circ$\\
\noalign{\smallskip}
\tableline
\noalign{\smallskip}
0.5 &   6.7 &  1.5  \\
1.0 &  24.4 &  5.7  \\
1.5 &  46.7 & 12.3  \\
2.0 &  67.5 & 20.6  \\
2.5 &  82.9 & 30.0  \\
3.0 &  92.1 & 40.0  \\
3.5 &  96.7 & 49.9  \\
4.0 &  98.7 & 59.2  \\
4.5 &  99.4 & 67.8  \\
5.0 &  99.7 & 75.4  \\
\noalign{\smallskip}
\tableline
\end{tabular}
}
\end{center}
\end{table}

\acknowledgements
The author is a Research Fellow of the Alexander von Humboldt Foundation.
The author would like to thank Ken Kellermann for fruitful discussions
and Leonid Petrov for providing Figure~\ref{f:skydist} and Table~\ref{t:calprob}.
This research has made use of NASA's Astrophysics Data System.


\end{document}